# Simulation on Buildup of Electron Cloud in Proton Circular Accelerator*

Liu Yu-Dong(刘瑜冬), Li Kai-Wei(李开玮)[1)]

(China Spallation Neutron Source (CSNS), Institute of High Energy Physics (IHEP), Chinese Academy of Sciences (CAS), Dongguan 523803, China)

**Abstract**

Electron cloud interaction with high energy positive beam are believed responsible for various undesirable effects such as vacuum degradation, collective beam instability and even beam loss in high power proton circular accelerator. An important uncertainty in predicting electron cloud instability lies in the detail processes on the generation and accumulation of the electron cloud. The simulation on the build-up of electron cloud is necessary to further studies on beam instability caused by electron cloud. China Spallation Neutron Source (CSNS) is the largest scientific project in building, whose accelerator complex includes two main parts: an H- linac and a rapid cycling synchrotron (RCS). The RCS accumulates the 80Mev proton beam and accelerates it to 1.6GeV with a repetition rate 25Hz. During the beam injection with lower energy, the emerging electron cloud may cause a serious instability and beam loss on the vacuum pipe. A simulation code has been developed to simulate the build-up, distribution and density of electron cloud in CSNS/RCS.

**Key words:** Electron Cloud, Instability, Electron Multipacting, Beam Loss

## I Introduction

In high intensity proton circular accelerator, the electron cloud effect has been considered as one of the main sources of beam instability, which can lead to the uncontrolled beam loss[1]. This electron proton instability has been observed and confirmed in many commission proton circular accelerators, such as LANL PSR [2], KEK Booster [3], CERN PS [4] and SNS in ORNL[5]. During proton beam operation, the main primary electrons produced by proton losses at the chamber surface are attracted and accelerated by the body part of bunch, then released at the bunch tail or bunch spacing. The secondary electrons are produced and amplified when these accelerated electrons hit the chamber wall. This secondary electron multipacting attributes to the main source of electron cloud.

---

\* Work supported by National Natural Science Foundation of China (11275221, 11175193)

1) Email: likw@ihep.ac.cn

A high intensity proton accelerator facility as neutron source proposed in past ten years and is being built in China, which is named China Spallation Neutron Source (CSNS) [5]. The facility is equipped with an H- linac and a rapid cycling synchrotron (RCS). RCS is designed to accumulate proton beam and accelerate it from 80MeV to 1.6GeV. The accumulated particles in the RCS can reach to $7.8 \times 10^{12}$ protons in a bunch with length about several tens of meters. Such high population and long bunches circulating in RCS may lead to denser electron cloud. A simulation code has been developed and benchmarked to ascertain detailed establishment of electron cloud in CSNS/RCS. The physical mechanism on primary electron production, its dynamics, secondary electron emission and electron accumulation is described in section II of this paper. By tracking dynamics of electrons with physical model in section II, a simulation code for obtaining distribution and density of electron cloud in proton circulated accelerator has been developed and benchmarked with other machines. With the beam parameters of RCS/CSNS, the process of electron cloud buildup is studied in different simulation conditions such as average beam loss factor, secondary electron yields, transverse and longitudinal beam profile, and beam intensity. Most of the simulations are performed in a field-free region mainly, and the contribution of the dipole magnetic field is also investigated. The comprehensive simulation on electron cloud in RCS/CSNS gives the quantitative relation between electron density and beam parameters which is meaningful to the construction of RCS/CSNS.

Table 1: The main parameters of the RCS/CSNS

| Parameters | Injection Phase | Extraction Phase |
|---|---|---|
| Circumference $C$ (m) | 227.92 | 227.92 |
| Energy $E$ (GeV) | 0.08 | 1.6 |
| Bunch Population $N_p$ ($10^{13}$) | 1.56 | 1.56 |
| Revolution Frequency $\omega$ (MHz) | 3.20 | 6.78 |
| Bunch length $\sigma_p$ (m) | 48.93 | 22.285 |
| Harmonic number $h$ | 2 | 2 |
| Beam transverse size $\sigma_x$, $\sigma_y$ (cm) | 1.5, 1.5 | 1.2, 1.2 |
| Pipe radius $b$ (cm) | 10 | 10 |

## II  Physical Model

The electron sources in proton circular accelerator may be classified into (1) electrons stripped at the injection region; (2) electrons produced by proton losses incident the vacuum chamber at grazing angles; (3) secondary electron emission process; and (4) electrons produced by residual gas ionization. The stripped electrons generated near the stripping foil, which can be absorbed by installing a special electron collector at the injection region [7]. The yield of ionization electrons is determined by the ionization cross section and the vacuum pressure in the beam chamber. Because of the lower vacuum pressure about $10^{-7}$ Pa and ionization cross section for CO and H2 approximately 1.3 MBar and 0.3 MBar, the ionization electrons can be neglected in the simulation. In the electron dynamic tracking for RCS/CSNS, the primary electrons produced by proton losses and secondary electrons are considered.

When primary electrons are emitted from the pipe wall, they have interaction with the proton beam and move towards the opposite wall, and then hit the wall after a period of transit. These striking electrons may produce secondary electrons if their energy is high enough. The fraction between the emission electrons from pipe surface and the total incident electrons is defined as secondary electrons yield (SEY). If SEY is above 1, the electrons multipacting will happen. Actually, the secondary electrons include two types: the elastic backscattered electrons and true secondary electrons. The yields of the true secondary electrons and elastic backscattered electrons can be expressed with formula (1) and (2) [8], respectively. In the formula, $\delta_{max}$ is the maximum secondary emission yield for perpendicular incidence, and $\varepsilon_{max}$ presents the incidence electron energy with maximum secondary emission yield $\delta_{max}$. Here, $E_p$ is the energy of the electron, $\theta$ the angle of the incidence electrons and $\hat{\delta}_e, \delta_{e,\infty}, \Delta \text{ and } E_e$ are the experiential fitting parameters decided by pipe material with $\hat{\delta}_e = 0.1, \delta_{e,\infty} = 0.02$ and $\Delta = E_e = 5eV$ [8].

$$\delta_{se}(E_p, \theta) = \delta_{max} \times \frac{1.44 \times (E_p / \varepsilon_{max})}{0.44 + (E_p / \varepsilon_{max})^{1.44}} \exp(0.5 \times (1 - \cos(\theta))) \qquad (1)$$

$$\delta_e(E_p) = \delta_{e,\infty} + \hat{\delta}_e - \delta_{e,\infty} \exp(\frac{-(E_p - E_e)^2}{2\Delta^2}) \qquad (2)$$

The results on SEY with and without elastic backscattered electrons are illustrated in figure 1. It is clear that the elastic portion have more influence in the lower energy region. In the simulation, these two types of secondary electrons are included.

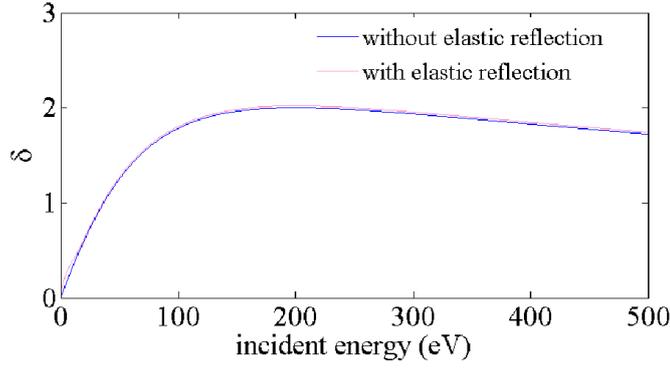

Figure 1. Secondary electron yield in different incident electron energy

In the magnetic-field-free region, the electrons move under the space-charge fields of the proton beam and between other electrons. For the long proton bunch, the longitudinal space-charge field can be neglected due to slow potential variations in longitudinal direction and the symmetry of longitudinal beam profile that traps the particles longitudinally. Therefore, the electrons mainly move under the beam transverse fields. For a cylindrical beam with uniform and gaussian transverse distribution, the space-charge fields are expressed respectively as

$$E_r(t) = \begin{cases} \dfrac{\lambda(t)}{4\pi\varepsilon_0}\dfrac{2}{r} & (r > \sigma_r), \\ \dfrac{\lambda(t)}{4\pi\varepsilon_0}\dfrac{2r}{\sigma_r^2} & (r < \sigma_r), \end{cases} \quad (3)$$

$$E_r(t) = \dfrac{\lambda(t)}{4\pi\varepsilon_0}[1-\exp(-\dfrac{r^2}{2\sigma_r^2})]\dfrac{2}{r} \quad (4)$$

where $\lambda(t)$ is the beam's line density, $\varepsilon_0$ known as the permittivity of vacuum, and $\sigma_r$ is the beam transverse size. In fact, the space-charge fields of any spatial particle distribution can be obtained by numerical evaluation with PIC methods, which meshes the particles in three dimensional grid and makes two dimensional FFT in the transverse grids for each longitudinal slice to compute the potential on every grid point. The electric field comparison between numerical solution and analytical solution to uniform and Gaussian distributed charge are shown in figure2. It is clear the numerical solution fits well to the analytical results.

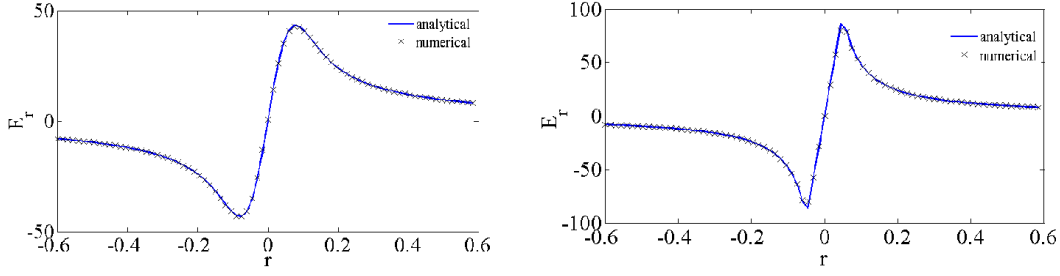

Figure 2. Electric field with different particle distribution (left: Gaussian; right: uniform)

During the passage of the proton bunch, the electrons experience the beam action and make a nearly periodic oscillation with a slow time dependence given by the beam line density $\lambda(t)$. Assuming the bunch is uniform in the longitudinal direction with bunch population $1.56 \times 10^{13}$ in CSNS/RCS, the oscillation frequency of the electron ranges from 20.23 MHz to 134.84 MHz.

In the simulation model, electron cloud formation is estimated by tracking the motion of electrons produced by the primary and the secondary electron emissions. The vacuum chamber is assumed to be a cylindrical perfectly conducting pipe. The primary electrons generated by lost protons hitting the vacuum chamber wall is $Y*P_{loss}$ per turn for the whole ring, where $Y$ is the effective electron yield per lost proton, and $P_{loss}$ is the proton loss rate per turn for the whole ring per beam proton. The lost-proton time distribution is proportional to the instantaneous bunch intensity. The electrons are then represented by macroparticles. The secondary electron mechanism adds to these a variable number of macroparticles, generated according to the SEY model mentioned above. The proton bunch is sliced along the longitudinal direction into equal-size steps and each slice has a local proton density $\lambda(t_i)$. Electrons are tracked step by step along the passage of the proton beam. The equation of motion for electrons is expressed by

$$\frac{d^2 x(t)}{dt^2} = -2\lambda_p(t_i) r_e c^2 F_G[x(t)] \tag{5}$$

Where $r_e$ is the classical electron radius, $F_G$ is the normalized electric force determined by equation (3) and (4). For proton bunch in CSNS/RCS, the longitudinal profile is sinusoidal function expressed as

$$\lambda_p(t_i) = \frac{\pi N_p}{2\sigma_p} \sin(\frac{\pi \upsilon t_i}{\sigma_p}) \tag{6}$$

Where $\sigma_p$ is the longitudinal length of theproton bunch and $\upsilon$ is bunch velocity.

## III  Simulation Results

According to the physical model described in section II, a simulation code was developed independently to understand the buildup of electron cloud in proton circular accelerator. After verifying the validity of simulation by benchmarking electron cloud density in J-PARC, LANL PSR and ORNL SNS, the buildup of the electron cloud in CSNS/RCS is simulated in different parameters, such as proton loss rate, SEY, beam intensity and bunch transverse size. For high intensity proton accelerator, the common requirement of loss power for maintenance is lower than 1 W/m. Assuming the beam loss happen in first several hundred turns and every lost proton can produce 100 electrons [9], the densities of electron cloud in series of $P_{loss}$ are shown in figure 3.

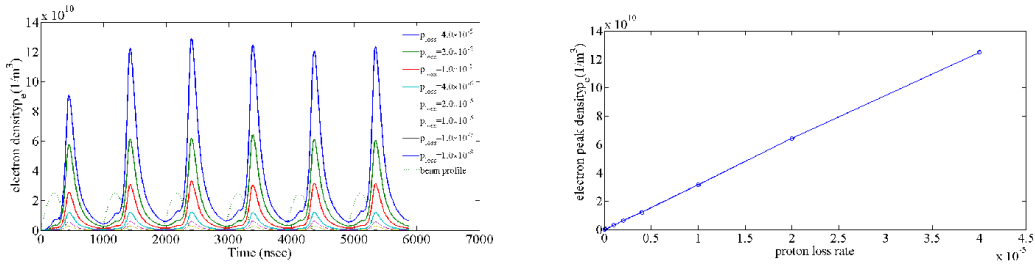

Figure 3. The densities of electron cloud in different proton loss rate

It is clear that the peak density of the electron cloud linearizes with the proton loss rate because of the thinner density of electron cloud whose line density $\lambda_e$ is only about 2% proton average line density $\lambda_p$. The transverse distribution of electrons during the passage of the bunch for CSNS/RCS 80 MeV at injection is shown in figure 4.

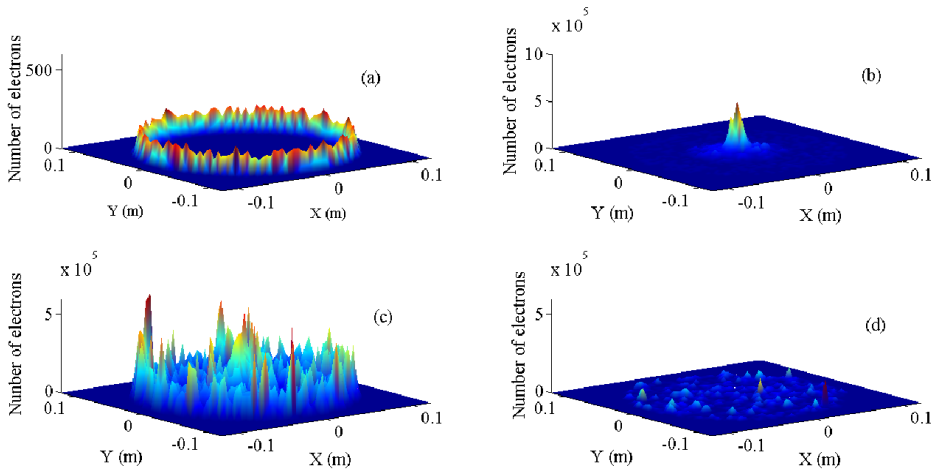

Figure 4. Transverse distribution of electron cloud during a bunch passage

(a: bunch head; b: bunch center; c: bunch tail; d: bunch gap)

It is clear the electrons are produced near the pipe wall and distributed widely at the start of the interaction with the bunch. Then these electrons are gathered to the beam vicinity and the cloud size is comparable to the beam size. After the bunch passage and without interaction, these accumulated electrons splash in the vacuum pipe. The last picture shows the electrons distribution before the next bunch coming with a considerable quantity of electrons remaining in the chamber. This quantity depends on the ring and beam parameters. During the passage of the next bunch, the same process will happen again.

The secondary electrons produced by primary electron incidence on the pipe wall will be main portion of the electron cloud when the SEY above 1.0. The TiN coating inner the vacuum can suppress the maximum secondary electron yield to 1.3. The electron cloud amplification in various $\delta_{max}$ is simulated for CSNS/RCS and the results are shown in figure 5. It is clear that the peak electron cloud density is sensitive to the maximum secondary electron yield. So the most efficient method to cure the electron cloud is surface treatment to reduce $\delta_{max}$ below 1.3.

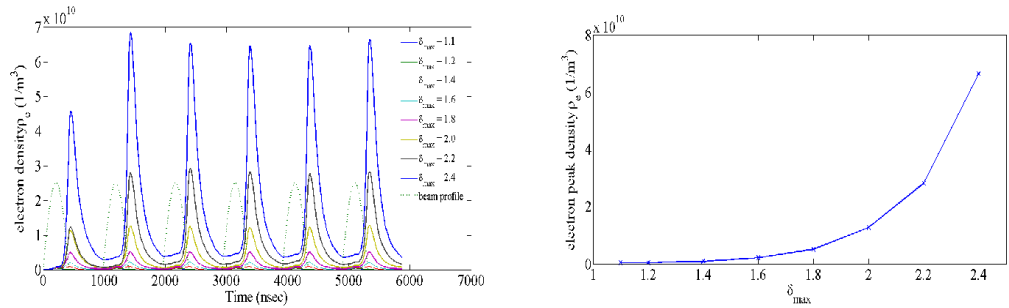

Figure 5. The electron cloud with various $\delta_{max}$

The energy of primary electrons is decided by the loss protons. In order to understand the relation between primary electron characteristics and electron cloud, the development of eletron cloud in different primary electrons energy is simulated and shown in figure 6. The secondary electron yield parameters, $\delta_{max}=2.0$ and $\varepsilon_{max}=200eV$ are adopted in the calculation. The electron cloud density in figure 6 varies as the SEY curve because of the electron multipacting dependence on the incident energy.

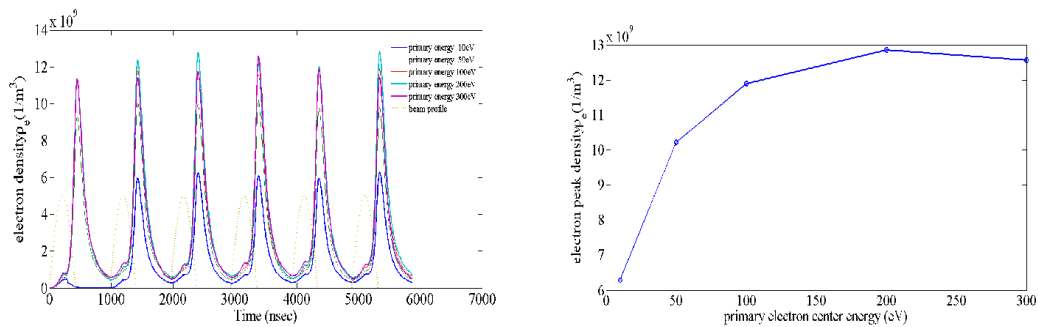

Figure 6. The electron cloud vs various primary electron energy

The beam transverse size also has much influence on the saturation density of the electron cloud because of the beam space-charge fields expressed in equation (3) and (4). The simulation on electron cloud under different beam transverse size is shown in figure 7. Initially, the narrow beam transverse size lead to a stronger electronic field and the electron energy gain exceeds the SEY parameter $\varepsilon_{max}$. With the beam size extension, the electron energy gain declines to the $\varepsilon_{max}$ and the electron multipacting is serious to attribute the highest cloud density. Broadening the beam size gradually, electron energy gain drops lower to the $\varepsilon_{max}$ and the cloud density descends correspondingly.

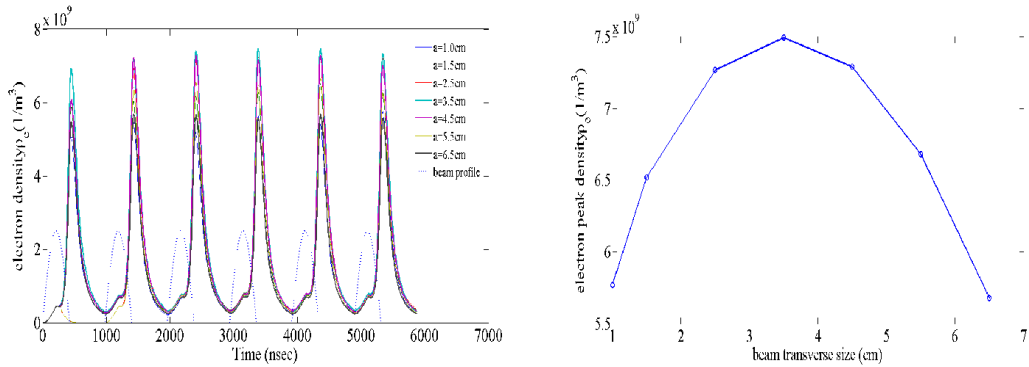

Figure 7. Electron cloud density vs beam transverse size

The electron cloud buildup with different beam intensity is plotted in figure. With the stronger beam intensity, the primary electrons produced by beam loss increase correspondingly and their energy gain is enhanced simultaneously. That reasons lead to the electron cloud density raise exponentially, shown in figure 8. When the beam intensity reaches to $4.0 \times 10^{13}$/bunch, the peak density of electrons cloud is more than $7.6 \times 10^{11}$/m$^3$.

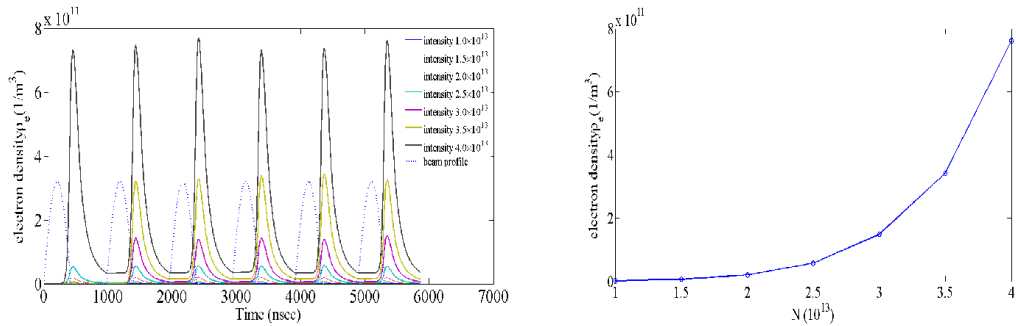

Figure 8. Electron cloud density vs different beam intensity

We now discuss the electron cloud buildup in magnetic field. The beam chamber is covered with various magnets as bending, quadrupole, and higher order magnets in actual rings. The field-free regions and bending magnets occupied the biggest part of the rings. In bending magnets with a strong dipole field, electrons undergo cyclotron motion with a small radius ( <1 mm) and at a high frequency ( > 10 GHz). The electron cloud buildup in dipole magnet is plotted in figure 9. Because of the cyclotron motion caused by the dipole field, considerable fraction of the electrons cannot approach the beam, therefore they don't get sufficient energies for the multipacting. So the dipole fields have the effect of suppressing the buildup of electron cloud.

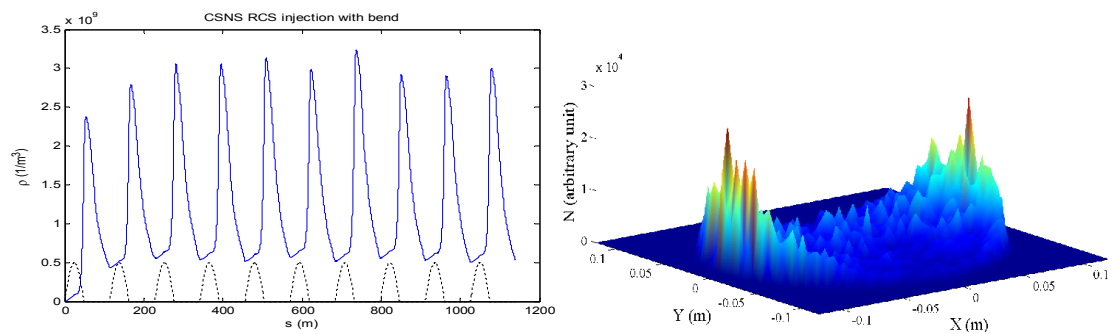

Figure 9. Electron cloud in dipole magnets

## IV   Conclusion

In instance of CSNS/RCS, the buildup of the electron cloud in circular proton accelerator was investigated in this paper. A computer simulation code is developed to study the accumulation process of electron cloud. In the simulation model, primary electrons appear due to proton loss on the wall of vacuum pipe and the secondary electron emission model is also included. The calculation proved the application on TiN coating to reduce the secondary yield is a very powerful cure for electron cloud. The investigation on the electron cloud buildup in various beam parameters such as beam intensity, transverse size and proton loss rate, is done in detail. The formation of the electron cloud in dipole field is also simulated to testify the suppression on electron multipacting. All these simulation will be meaningful to understand the interaction between electrons and high intensity proton beam in circular accelerators.